\documentstyle[aps,preprint]{revtex}
\begin{document}
\newcommand{\r}{\vec{r}\!\:}           
\newcommand{\p}{\vec{p}\!\;}           
\newcommand{\ld}{\vec{\bf l}\,}      
\newcommand{\Ld}{\vec{\bf L}}        
\newcommand{\sd}{\vec{\bf s}}        
\newcommand{\Sd}{\vec{\bf S}}        
\newcommand{\jd}{\vec{\bf j}}        
\newcommand{\Jd}{\vec{\bf J}}        
\newcommand{\vnabla}{\vec{\nabla}}   
%
\newcommand{\backsix}{\! \! \! \! \! \!}
\newcommand{\bbb}{\! \! \! }
\newcommand{\bb}{\! \! }
\newcommand{\hp}{\hspace{0.5cm} .}
\newcommand{\hc}{\hspace{0.5cm} ,}
\newcommand{\abinitio}{\glqq ab initio\grqq \ }
\newcommand{\GAUSS}{ GAUSSIAN\,90 und GAUSSIAN\,92 }
\newcommand{\BK}{ Bona\v{c}i\'{c}-Kouteck\'{y} }

\newcommand{\hhn}{$ {\rm H}_3^+ {\rm (H}_2{\rm )}_n$ }
\newcommand{\hhnX}{$ {\rm H}_3^+ {\rm (H}_2{\rm )}_{n-1}$ }
\newcommand{\Heins}{${\rm H}_1$ }
\newcommand{\Hzwei}{${\rm H}_2$ }
\newcommand{\Hdrei}{${\rm H}_3^+$ }
\newcommand{\Hfuenf}{${\rm H}_3^+ {\rm (H}_2{\rm )}$ }
\newcommand{\Hsieben}{${\rm H}_3^+ {\rm (H}_2{\rm )}_2$ }
\newcommand{\Hneun}{${\rm H}_3^+ {\rm (H}_2{\rm )}_3$ }
\newcommand{\Helf}{$ {\rm H}_3^+ {\rm (H}_2{\rm )}_4$ }
\newcommand{\Hdreizehn}{$ {\rm H}_3^+ {\rm (H}_2{\rm )}_5$ }
\newcommand{\Hfuenfzehn}{$ {\rm H}_3^+ {\rm (H}_2{\rm )}_6$ }
\newcommand{\Hsiebzehn}{$ {\rm H}_3^+ {\rm (H}_2{\rm )}_7$ }
\newcommand{\Hneunzehn}{$ {\rm H}_3^+ {\rm (H}_2{\rm )}_8$ }
\newcommand{\Heinundzwanzig}{${\rm H}_3^+ {\rm (H}_2{\rm )}_9$ }
\newcommand{\Hdreiundzwanzig}{${\rm H}_3^+ {\rm (H}_2{\rm )}_{10}$ }
\newcommand{\Hfuenfundzwanzig}{${\rm H}_3^+ {\rm (H}_2{\rm )}_{11}$ }
\newcommand{\Heinunddreissig}{${\rm H}_3^+ {\rm (H}_2{\rm )}_{14}$ }
%
%

\title{
Structures and Stabilities of \hhn Clusters (n=1-11) }
\author{
   B.\ Diekmann, P.\ Borrmann, E.R.\ Hilf }
\address{Department of Physics, Carl v. Ossietzky University Oldenburg,
   D-26111 Oldenburg, Germany}
\maketitle
\begin{abstract}
Geometries and energies for \hhn  clusters (n = 0, ..., 11)
have been calculated
using standard "ab initio" methods. Up to clusters with  n = 6,
four different Pople basis sets (DZ, TZ, TZP) have been used in the
calculations.
For larger cluster sizes, the calculations have been carried out with
one basis set (DZ) using the HF/CISD method.
We discuss here only the nice counterplay of polarisation effects between  the
central \Hdrei ion and the adsorbed \Hzwei molecules, which naturally
governs both the geometric structure and the energy of the clusters. \\
{\bf Please regard: This article is also availible in the WWW under the
URL: \\
{\small
http://www.physik.uni-oldenburg.de/Docs/documents/UOL-THEO3-94-5/cont.html}
 }
\end{abstract}
%
%
\section{Introduction}
\Hdrei molecules have been first experimentally verified in
1912 \cite{JThomson_1912},
Dawson and Tickner have identified \Hfuenf
clusters by mass spectometry in 1962 \cite{Dawson_1962} .
In a recent work of Kirchner and Bowers on this topic
\cite{Kirchner_1983} \cite{Kirchner_1986}
the dynamics of metastable \Hfuenf
fragmentations are studied. Another
experimental work of Hiraoka and Mori
\cite{Hiraoka} deals with the stability of \hhn clusters.

Many theoretical studies of small \hhn clusters ( n $\le$ 3 )
have been published in the last years. Yamaguchi, Gaw,
Remington and  Schaefer \cite{Yamaguchi_1987} have made an
intensive study of \Hfuenf  with the Pople Double
Zeta plus Polarization Basis (DZP),
\Hsieben, \Hneun, \Helf and \Hdreizehn
have been calculated with the
Triple Zeta plus Polarization Basis (TZP) by
Farizon, Farizon-Mazuy, Castro Faria and Chermette
\cite{Farizon_03.1991}, \cite{Farizon_07.1991}
and \cite{Farizon_01.1992}.
To our knowledge, this ab initio calculation is the first calculation
for \hhn clusters for $n$ larger than 5. We will study here clusters
up to $n = 11$ and compare the lighter ones the results of Farizon
et al. The results will be discussed in terms of geometrical
izomerization.
%
%
\section{Numerical procedure}
The geometric optimization of the hydrogen clusters was implemented here by the
BERNY-algorithm \cite{BernySchlegel}.
In the optimization no restrictions
in bond-lengths and bond-angles have been adopted. The geometric
construction of the clusters is done sequentially,
starting with an optimized
\Hdrei structure with a randomly placed \Hzwei  molecule added at a
distance of about 3.5 {\AA} from the \Hdrei . In a first run the
position of the new \Hzwei has been optimized, in a subsequent calculation
all geometric variables have been optimized. The optimizations of
the larger clusters are performed analogously with each new \Hzwei molecule
placed randomly at a distance of about 3.5 {\AA} from the
last added \Hzwei molecule. \\
For the electronic configuration we selected the CISD/DZ and the CISD/TZP model
for two sets of calculations.
With regard to the electron-correlation effects in \Hzwei
and \Hdrei , CI calculations should be
used \cite{Bernd_Diplomarbeit}. \\
For our intention to make predictions for larger clusters,
we compare a model on a relatively low theoretical level
but suited to be used for larger clusters, with a more
sophisticated ansatz, both applied to small clusters.
After comparing these results it should be
possible to extrapolate properties of larger clusters
using the simpler ansatz.  \\
For our extensive numerical calculations we adopted the
GAUSSIAN 90 and GAUSSAIN 92 code \cite{GaussInc}
on a S400/4 supercomputer \cite{Fuijutsu_HA}, which has a
peak performance of 5 GFlops.
Calculations of such relatively large
systems with such large basis sets were only possible by use of an extra
1 GB ram-memory besides the main memory of up to 1 GB. The calculations took
all in all about 200 cpu hours. \\
The procedure described above has been tested by application to \Hdrei
and \Hfuenf clusters, which have been investigated thoroughly before. \\
The results of the calculations of the
${\rm H}_3^+ {\rm (H}_2{\rm )}$  cluster are
in good aggreement with the results of {\sc Farizon} et
al. \cite{Farizon_03.1991} and with other theoretical work.

%
%
\section{Results and Discussion}
The theoretical calculations, which include correlation
effects of the electrons, reduce the energies of the
systems.
Regarding the electrostatic properties of the $ {\rm H}_2 $
molecules, it is nessesary to use polarization functions
in these methods in geometric calculations.

Fig. 1 and Fig. 2 display the electron density in \Hfuenf calculated
with the TZP basis from different views.
Apparently the \Hzwei ist most tightly bound when closest to one of
the hydrogens of \Hdrei pushing the negative charges to the other
two hydrogens in the \Hdrei. \\
The more subtle effect of the orientation
of the \Hzwei is predicted by the TZP calculations to be perpendicular to the
\Hdrei-plane. Thus the electrons of \Hzwei are most distant
from the electrons  within the \Hdrei.\\
With the  less sophisticated basis set DZ (two instead of
six basis functions per electron) geometric
optimization yields a quite different result. The polarization is
less pronounced and incomplete, yielding a completely planar structure.
The electron  charges in the \Hdrei are almost equidistributed
over the three protons. Interestingly the electron density between
the \Hzwei and the \Hdrei comes out too low, resulting in a
reduction of the binding energy of as much as 0.05 a.u.\ .\\
Understanding what
happens in this smallest cluster makes it easy to understand
the larger structures, too.\\
The \Hsieben and \Hneun clusters are built by the same structure
principle, while the charge distribution is more amusing (the corresponding
geometry structures are shown in figure 4 and figure 5). A nice
example of charge frustration occurs in the \Hdrei, which means
that the charges are being repelled from all \Hzwei symmetrically
and assemble in the middle of the \Hdrei triangle. \\
A first geometric and energetic shell is filled with these three
molecules as is also indicated by the differences in the energies,
which are give in figure \ref{stabilities} .

For a study of the relative stability of the clusters we define a value
which reflects that the \hhn clusters are a composition of
\Hzwei  and \Hdrei molecules
\begin{equation}
 l := \frac{ E_{{\rm H}_n^+} }{
 E_{{\rm H}_3^+} + \frac{n-3}{2} E_{{\rm H}_2} }  \; ,
\end{equation}
as a measure of the relative stability of the clusters.
The relative stability reduces while the cluster size
increases.

Magic numbers using these  defined $l$-values may be inferred
for $ n = 3, 5 $ and
$ 9 $ using DZ.
The relative stability of \Hdreizehn
has been predicted by
Hirao and Yamabe \cite{Hirao_Yamabe_1983} in a
geometric consideration.
In the experimental work of  Hiraoka \cite{Hiraoka}
the \Hfuenfzehn seems to be
more stable than its neighbours.
Hiraoka related this to
a total planar geometry of \Hfuenfzehn . We could not confirm this
and found no planar structure. \\
The magic numbers of $ n = 3,5,9 $, see figure 7, demonstrate the
packing of the \Hzwei, where three of them form the innermost shell,
while the next shell consists of four \Hzwei.

%
%
%
In the calculations the model of the positive \Hdrei  center
of the cluster with neutal \Hzwei molecules has been verified. \\
A property of the geometry of the clusters are the interatomic distances of
the \Hzwei molecules. This effect depends directly on the distance from the
positive \Hdrei center of the cluster perturbated by the other \Hzwei as shown
in figure 6. The interatomic distances of the \Hzwei molecules
decrease with  increasing distance of those molecules
from the \Hdrei center. This
is another nice polarization effect which can be understood in the following
manner. The positive central charge attracts the electrons of the \Hzwei
reducing  the screening of the protons of the  \Hzwei which in turn
enlarges the repulsion between them.


We have calculated \hhn clusters up to $ n = 11 $, the largest \hhn cluster
calculated so far. We observe two glooming obstacles:
The optimization reveals more and more geometric isomers (second
minima in the energy) with more and more shallow barriers. \\
In addition the complexity of the basis set necessary
for a reliable calculation should increase exponentially
which is prevented by
the computational resources. \\
Finally the kinetic energies of the atoms contribute to the total energy
of the clusters substantially and are not taken into account in the
codes used here.

\newpage
\begin{figure}
\caption{Contour plot of the electron density of \Hfuenf
  with TZP basis taken in the plane of \Hdrei.}
\end{figure}
\begin{figure}
\caption{Contour plot of the electron density of \Hfuenf
 with TZP basis taken perpendicular to the plane of \Hdrei.}
\end{figure}
\begin{figure}
\caption{Contour plot of the electron density of \Hfuenf
 with DZ basis taken perpendicular to the plane of \Hdrei.}
\end{figure}
\begin{figure}
\caption{Three dimensional view of the \Hneun structure.}
 \end{figure}
\begin{figure}
\caption{Three dimensional view of the \Hneun structure.}
\end{figure}

\begin{figure}
\caption{\hhn Geometric properties: interatomic distances of \Hzwei as a
         function of the cluster size.}
\end{figure}
\begin{figure}
\caption{\hhn Stabilities: l-value of \hhn clusters as
          a function of the cluster size.
         \label{stabilities} }
\end{figure}

\begin{figure}
\caption{Three dimensional view of the \Hdreizehn structure.}
\end{figure}
\begin{figure}
\caption{Three dimensional view of the \Hsiebzehn structure.}
\end{figure}

\begin{table}
\caption{Energies (a.u.) of \hhn cluster calculations}
\begin{tabular}{llll}
                 &  HF/TZP    & RCISD/DZ  & RCISD/TZP  \\
\Hfuenf          &            & -2.46232  & -2.51659    \cr
\Hsieben         &            & -3.61718  & -3.68837    \cr
\Hneun           &            & -4.77087  & -4.85857    \cr
\Helf            & -5.84204   & -5.91698  & -6.02430    \cr
\Hdreizehn       &            & -7.06959  & -7.18842    \cr
\Hfuenfzehn      & -8.10608   & -8.21822  & -8.34985    \cr
\Hsiebzehn       &            & -9.36654  &             \cr
\Hneunzehn       & -10.37204  &-10.51389  &             \cr
\Heinundzwanzig  &            &-11.66103  &             \cr
\Hdreiundzwanzig &            &-12.80764  &             \cr
\Hfuenfundzwanzig&            &-13.95374  &             \cr
\end{tabular}
\end{table}
\end{document}